January, 2020

# The 5–CB Algebra and Fused $SU(2)$ Lattice Models

Vladimir Belavin[a,b,c], Doron Gepner[d]

[a] *Physics Department, Ariel University, Ariel 40700, Israel*

[b] *I.E. Tamm Department of Theoretical Physics, P.N. Lebedev Physical Institute, Leninsky av. 53, 11991 Moscow, Russia*

[c] *Department of Quantum Physics, Institute for Information Transmission Problems, Bolshoy Karetny per. 19, 127994 Moscow, Russia*

[d] *Department of Particle Physics and Astrophysics, Weizmann Institute, Rehovot 76100, Israel*

## ABSTRACT

We study the fused $SU(2)$ models put forward by Date et al., that are a series of models with arbitrary number of blocks, which is the degree of the polynomial equation obeyed by the Boltzmann weights. We demonstrate by a direct calculation that a version of BMW (Birman–Murakami–Wenzl) algebra is obeyed by five, six and seven blocks models, conjecturing that it is part of the algebra valid for any model with more than two blocks. To establish this conjecture, we assume that a certain ansatz holds for the baxterization of the models. We use the Yang–Baxter equation to describe explicitly the algebra for five blocks, obtaining 19 additional non–trivial relations. We name this algebra 5–CB (Conformal Braiding) algebra. Our method can be utilized to describe the algebra for any solvable model of this type and for any number of blocks.

## 1. Introduction.

Solvable lattice models in two dimensions are an excellent playground to test such ideas as phase transition, universality and integrability. For a review see the book [1]. Here, we will concentrate on Interaction Round the Face (IRF) solvable lattice models. Seminal examples of such models are the Andrews, Baxter and Forrester models [2], which generalizes the Hard Hexagon model and the Ising model [1]. Other important examples are the height models of the Kyoto group [3, 4]. For a review of IRF models see [5], and references therein.

The algebraic structure of such models was investigated since the work of Temperley and Lieb [6]. The algebra was essential in the solution of the models as well as in applications such as knot theory. For a review of the latter see, e.g. the book [7].

In ref. [8], an ansatz for the baxterization was given, which will be utilized in the main body of this paper. We believe that this ansatz holds for all the known solvable IRF lattice models and we assume that this ansatz is correct for the models that we discuss. This ansatz agrees with the two blocks baxterization found by Jones [9], and the results of Jimbo and Pasquier in the cases of $SU(2)$ IRF models [10, 11]. Also, in ref. [8], it was conjectured that an algebra called $n$–CB algebra, where $n$ is the degree of the polynomial equation obeyed by the Boltzmann weights, holds. In subsequent papers [12, 13, 14], we studied the three and four blocks cases and their associated algebras. We used a combination of the Yang–Baxter equation and the ansatz for Baxterization described in ref. [8], to deduce the underlying algebras. Thus, we gave a full description of the 3–CB as the Birman–Murakami–Wenzl (BMW) algebra [15, 16] and the 4–CB algebra as a BMW algebra with a different skein relation and one additional relation.

In this paper we study the $n$–CB algebra with five blocks or more, $n \geq 5$, and especially the 5–CB algebra. First, we use the fused $SU(2)$ $m \times m$ models of Date et al. [17]. These are height models based on the algebra $SU(2)$ which have $m+1$ blocks for every integer $m \geq 1$. We use these models as a test grounds for the $n$–CB algebra. We show, by explicit calculation, that these models obey the BMW algebra, with a different skein relation, for $n = 5, 6, 7$. We thus conjecture that the BMW algebra is obeyed for any model with three or more blocks.

Further, we use our ansatz combined with the Yang–Baxter equation along with the BMW algebra, to give the full relations of the 5–CB algebra for the $SU(2)$ models. We find 19 additional relations. Our method is general and can be used to give the $n$–CB algebra for any model and any number of blocks.

## 2. IRF fusion models.

The IRF models are defined on a two dimensional square lattice. On each site sits a variable which we take to be one of the primary fields of some conformal field theory (CFT) $\mathcal{O}$. We fix some primary field of the CFT $\mathcal{O}$ which we denote by $h$. We further assume that the field $h$ is real, $h = \bar{h}$.

The partition function is given by

$$Z = \sum_{\text{configurations}} \prod_{\text{plaquettes}} \omega \left( \begin{matrix} a & b \\ c & d \end{matrix} \middle| u \right), \tag{2.1}$$

where $a, b, c, d$ are some primary fields sitting on the vertices of the plaquettes and $u$ is the spectral parameter. Here $\omega$ is some Boltzann weight, to be specifIed.

We assume that the Boltzmann weight $\omega$ vanishes unless the fields around a plaquette obey the admissibility condition which is,

$$f_{ah}^{b} > 0, \quad f_{ch}^{d} > 0, \quad f_{ah}^{c} > 0, \quad f_{bh}^{d} > 0, \tag{2.2}$$

where $f_{ah}^{b}$ is the fusion coefficient of the conformal field theory $\mathcal{O}$, and $h$ is a fixed primary field which we assume to be real. We call this model IRF$(\mathcal{O}, h)$.

We find it convenient to use an index free notation for the Boltzmann weights by defining,

$$\langle a_1, a_2, \ldots, a_n | X_i(u) | b_1, b_2, \ldots, b_n \rangle = \omega \begin{pmatrix} a_{i-1} & a_i \\ b_i & a_{i+1} \end{pmatrix} \!\! \Bigg| u \Bigg) \prod_{j \neq i} \delta_{a_j, b_j}. \tag{2.3}$$

We assume that the model obeys the Yang Baxter equation,

$$X_i(u) X_{i+1}(u+v) X_i(v) = X_{i+1}(v) X_i(u+v) X_{i+1}(u), \tag{2.4}$$

which is the key to the solvability of the model, by implying that the transfer matrices commute for different values of the spectral parameters.

The fusion product of the primary field $h$ is given by

$$h \cdot h = \sum_{i=0}^{n-1} \psi_i, \tag{2.5}$$

where $\psi_i$ are some primary fields, $\psi_0 = 1$, the unit operator and $n$ is some integer. We call the theory an $n$ block theory. The order of the fields in eq. (2.5) is crucial and the YBE, eq. (2.4), is obeyed only for one particular order. At the present, this order needs to be established one model at a time. A general rule is that $\psi_{i+1}$ is contained in the product of the adjoint representation with $\psi_i$ (for quantum group models), but the complete implementation of this rule is unclear. In this paper, we will be interested, mainly, in the five block theories. We define the crossing parameters as

$$\zeta_i = \pi(\Delta_{i+1} - \Delta_i)/2, \tag{2.6}$$

where $\Delta_i$ is the dimension of the field $\psi_i$ and $i = 0, 1, 2, \ldots, n-2$.

We define the operators $G_j$ and $G_j^{-1}$ as the limits of the Boltzmann weights,

$$G_j = X_j e^{-i\zeta_0(n-1)/2} \prod_{r=0}^{n-2} 2\sin(\zeta_r), \tag{2.7}$$

$$G_j^{-1} = X_j^t e^{i\zeta_0(n-1)/2} \prod_{r=0}^{n-2} 2\sin(\zeta_r), \tag{2.8}$$

where

$$X_j = \lim_{u\to i\infty} e^{i(n-1)u} X_j(u), \qquad X_j^t = \lim_{u\to -i\infty} e^{-i(n-1)u} X_j(u). \tag{2.9}$$

With this definition $G_i^{-1}$ is the inverse of $G_i$, $G_i G_i^{-1} = 1_i$.

It can be seen that $X_i$ obeys an $n$th order polynomial equation,

$$\prod_{r=0}^{n-1} (X_i - \lambda_r) = 0, \tag{2.10}$$

where the eigenvalues, $\lambda_i$ are given in terms of the conformal dimensions. We define projection operators by,

$$P_i^a = \prod_{\substack{p=0\\ p\neq a}}^{n-1} \left[\frac{X_i - \lambda_p}{\lambda_a - \lambda_p}\right]. \tag{2.11}$$

The projection operators obey the relations,

$$\sum_{a=0}^{n-1} P_i^a = 1_i, \qquad P_i^a P_i^b = \delta_{ab} P_i^b, \qquad \sum_{a=0}^{n-1} \lambda_a P_i^a = X_i. \tag{2.12}$$

We our now in position to state our ansatz for the trigonometric solution of the Yang–Baxter equation [8]. This is expressed as

$$X_i(u) = \sum_{a=0}^{n-1} f_a(u) P_i^a, \tag{2.13}$$

where the functions $f_a(u)$ are

$$f_a(u) = \left[\prod_{r=1}^{a} \sin(\zeta_{r-1} - u)\right] \left[\prod_{r=a+1}^{n-1} \sin(\zeta_{r-1} + u)\right] \Big/ \left[\prod_{r=1}^{n-1} \sin(\zeta_{r-1})\right]. \tag{2.14}$$

## 3. Fused $SU(2)$ IRF lattice models.

We turn our attention now to the fused $SU(2)$ IRF lattice models. Their Boltzmann weights were given by Date et al. [17]. In the language of section (2), these can be thought of as the models IRF($SU(2)_k$,$n/2$), namely the conformal field theory $\mathcal{O}$ is an $su(2)$ WZW model at level $k$ and the field $h$ is taken as the isospin $n/2$ field. For a review of conformal field theory, see, e.g., [18], and references therein. The theory is an $n+1$ block theory as the product of $h$ with itself is given by

$$h \cdot h = [n/2] \cdot [n/2] = \sum_{r=0}^{n} [r]. \tag{3.1}$$

The dimension of the field $[j]$ in an $SU(2)_k$ WZW model is given by

$$\Delta_j = \frac{j(j+1)}{k+2}. \tag{3.2}$$

Thus the crossing parameters, eq. (2.6), are given by

$$\zeta_j = \pi(\Delta_{j+1} - \Delta_j)/2 = \frac{\pi(j+1)}{k+2}, \qquad \text{for } j = 0, 1, \ldots, n-2. \tag{3.3}$$

Let us give now the Boltzmann weights following ref. [17]. We define

$$s[x] = \sin(\lambda x), \tag{3.4}$$

$$[x]_m = s[x]s[x-1]\dots s[x-m+1], \tag{3.5}$$

$$\begin{bmatrix} x \\ m \end{bmatrix} = [x]_m/[m]_m. \tag{3.6}$$

where $\lambda = \zeta_0 = \pi/(k+2)$.

We denote the Boltzmann weights by

$$\omega \left( \begin{matrix} a & b \\ c & d \end{matrix} \middle| u \right)_{p,q}, \tag{3.7}$$

as the Boltzman weight at the site variables $a$, $b$, $c$, $d$ at the sites around the plaquette. We denote the site variables by the dimensions of the representations, $a = 2m+1$, where $m$ is the isospin of the representation, etc. Here $p$ and $q$ are the fused horizontal $p$ plaquettes and vertical $q$ plaquettes. Our case of IRF$(SU(2)_k, [n/2])$ corresponds to $p = q = n$. We impose the 'unrestricted' admissibility condition, which corresponds to generic $k$ (where the restricted corresponds to integer positive $k$). These imply that the Boltzmann weights vanish unless,

$$(a-b+n)/2 \in \{0, 1, \dots, n\}, \tag{3.8}$$

where $n = p = q$ and $a$ and $b$ are the heights of any two adjacent sites. In the restricted models, where $k$ is an integer, we impose by the fusion rules,

$$a \in \{1, 2, \dots, k+1\}, \qquad n < a+b < 2k+4-n, \tag{3.9}$$

where $a$ is any height and $a$ and $b$ are any two adjacent sites.

We have the following formulas for the Boltzmann weights from Date et al. [17],

$$\omega\begin{pmatrix} l & l+2m-p \\ l+2r-q & l+2m-p+q \end{pmatrix}\!\Bigg|u\Bigg)_{p,q} = \frac{\begin{bmatrix} p-m \\ q-r \end{bmatrix}\begin{bmatrix} l+m+r-p-1 \\ r \end{bmatrix}\begin{bmatrix} m+u \\ r \end{bmatrix}\begin{bmatrix} l+m+u \\ q-r \end{bmatrix}}{\begin{bmatrix} l+r \\ q-r \end{bmatrix}\begin{bmatrix} l+2r-q-1 \\ r \end{bmatrix}}, \tag{3.10}$$

$$\omega\begin{pmatrix} l & l+2m-p \\ l+2r-q & l+2m-p-q \end{pmatrix}\!\Bigg|u\Bigg)_{p,q} = \frac{\begin{bmatrix} m \\ r \end{bmatrix}\begin{bmatrix} l+m \\ q-r \end{bmatrix}\begin{bmatrix} p-m+u \\ q-r \end{bmatrix}\begin{bmatrix} l+m-p+r-1-u \\ r \end{bmatrix}}{\begin{bmatrix} l+r \\ q-r \end{bmatrix}\begin{bmatrix} l+2r-q-1 \\ r \end{bmatrix}}. \tag{3.11}$$

Here $l$, $m$ and $r$ are integers.

Using these weights, the general Boltzmann weights are given by the recursion formula,

$$\omega\left(\begin{matrix} l & l' \\ l+2r-q & l'+2s-q \end{matrix}\,\middle|\,u\right)_{p,q}\begin{bmatrix} q \\ s \end{bmatrix} = \sum_{j=\max(0,r+s-q)}^{\min(r,s)} \omega\left(\begin{matrix} l & l' \\ l+2j-s & l'+s \end{matrix}\,\middle|\,u-q+s\right)_{p,s}$$
$$\times\omega\left(\begin{matrix} l+2j-s & l'+s \\ l+2r-q & l'+2s-q \end{matrix}\,\middle|\,u\right)_{p,q-s}. \tag{3.12}$$

To be consistent with the previous section, we find it convenient to substitute $u \to -u$ in the Boltzmann weights $\omega$.

## 4. BMW$'$ algebra.

We already defined the elements $G_i$ and $G_i^{-1}$, eqs. (2.7,2.8). We also identify

$$E_i = X_i(\lambda), \qquad 1_i = X_i(0), \tag{4.1}$$

where the crossing parameter is $\lambda = \zeta_0$. From the crossing relation and from the ansatz eqs. (2.13,2.14), we prove the Temperley–Lieb algebra for any number of blocks,

$$E_i E_{i\pm1} E_i, \qquad E_i^2 = bE_i, \qquad b = \prod_{r=0}^{n-2} \frac{\sin(\lambda+\zeta_r)}{\sin(\zeta_r)}, \tag{4.2}$$

and

$$E_i E_j = E_j E_i \quad \text{if } |i-j| \geq 2. \tag{4.3}$$

From the Yang–Baxter relation, we have the braiding equation,

$$G_i G_{i+1} G_i = G_{i+1} G_i G_{i+1}, \qquad G_i G_j = G_j G_i \ \text{ if } |i-j| \geq 2. \tag{4.4}$$

Another relation, which follows from the definition of the face transfer matrix, eqs. (2.7-2.9), is

$$G_i E_i = E_i G_i = l^{-1} E_i, \quad \text{where } \ l = i^{n-1} \exp\left[ i \left( (n-1)\lambda/2 + \sum_{r=0}^{n-2} \zeta_r \right) \right]. \tag{4.5}$$

From the skein relation, which will be discussed below, follows the relation,

$$G_{i\pm1} G_i E_{i\pm1} = E_i G_{i\pm1} G_i. \tag{4.6}$$

The above relations are proved to hold for any number of blocks greater than three, $n \geq 3$. To these relations we add a version of the Birman–Murakami–Wenzl (BMW) algebra [15, 16], which we conjecture to hold for any number of blocks greater or equal three, $n \geq 3$. The algebra, unlike BMW algebra, have a different skein relation. We call this algebra BMW$'$. The relations are,

$$G_{i\pm1} G_i E_{i\pm1} = E_i E_{i\pm1}, \qquad G_{i\pm1} E_i G_{i\pm1} = G_i^{-1} E_{i\pm1} G_i^{-1},$$

$$G_{i\pm1} E_i E_{i\pm1} = G_i^{-1} E_{i\pm1}, \qquad E_{i\pm1} E_i G_{i\pm1} = E_{i\pm1} G_i^{-1},$$

$$E_i G_{i\pm1} E_i = l E_i, \qquad E_i G_{i\pm1}^{-1} E_i = l^{-1} E_i. \tag{4.7}$$

The BMW$'$ algebra is a sub–algebra of the full algebra of the $n$ block model, which we call $n$–CB (Conformal Braiding) algebra. For three blocks and four blocks we verified the BMW$'$ algebra in the previous works [12, 13, 14]. We checked this conjecture for higher number of blocks, we use the fused $su(2)$ $m \times m$ models described in section (3), which are $n = m + 1$ blocks theories. We find it convenient to check it for the unrestricted models for a general $q = \exp(i\pi/(k+2))$. We find that this algebra is obeyed, indeed, numerically, for $n = 5, 6, 7$, with various values for the heights.

## 5. $n = 5$ CB case.

In this section we describe the derivation of the 5-CB relations, which is based on the Yang Baxter equation (YBE), (2.4), and on the ansatz (2.13,2.14). For $n = 5$ case we introduce parameters $s_k = e^{i\zeta_k}$, where $k = 0, ..., 3$. Using the relations involving the projector operators, (2.12), we can write the general Boltzmann weight, $X_i(u)$, as linear combination of the projectors $P_i^a$, with $a = 0, ..., 3$ and the identity operator. Using the defining relations, eqs. (2.7), (2.8), (4.1), for the operators $G_i, G_i^{-1}, E_i$, which are generators of the desired 5-CB algebra, we can write these generators as linear combinations of the projectors. In order to express the projectors in terms of the generators we introduce one more generator, $G_i^2$, which is written as linear combination of the projector using second relation in (2.12). The obtained linear problem gives

$$
\begin{aligned}
P_i^0 &= \frac{s_0^4 \left(s_1^2-1\right)\left(s_2^2-1\right)\left(s_3^2-1\right) g(3,i)}{\left(s_0^2+1\right)\left(s_0^2 s_1^2-1\right)\left(s_0^2 s_2^2-1\right)\left(s_0^2 s_3^2-1\right)},\\
P_i^1 &= -\frac{s_2^3 s_3 s_1^5 g(2,i)}{s_0\left(s_1^2+1\right)\left(s_1^2 s_2^2-1\right)\left(s_1^2 s_2^2 s_3^2+1\right)} + \frac{s_2^2\left(s_2^2 s_3^2 - s_3^2+1\right) s_1^4 g(5,i)}{\left(s_1^2+1\right)\left(s_1^2 s_2^2-1\right)\left(s_1^2 s_2^2 s_3^2+1\right)} +\\
&+ \frac{s_0 s_2 s_3 \left(s_3^2 s_2^2 - s_2^2+1\right) s_1^3 g(1,i)}{\left(s_1^2+1\right)\left(s_1^2 s_2^2-1\right)\left(s_1^2 s_2^2 s_3^2+1\right)} - \frac{s_0^2 s_2^2 s_3^2 s_1^2 g(4,i)}{\left(s_1^2+1\right)\left(s_1^2 s_2^2-1\right)\left(s_1^2 s_2^2 s_3^2+1\right)} +\\
&+ \frac{\left(s_1^2-1\right)\left(s_2^2-1\right)\left(s_0^2 s_1^2 s_2^2+1\right)\left(s_3^2-1\right)\left(s_0^2 s_1^2 s_2^2 s_3^2-1\right) g(3,i)}{\left(s_0^2+1\right)\left(s_1^2+1\right)\left(s_0^2 s_2^2-1\right)\left(s_1^2 s_2^2-1\right)\left(s_0^2 s_3^2-1\right)\left(s_1^2 s_2^2 s_3^2+1\right)},\\
P_i^2 &= \frac{s_1 s_3 s_2^3 g(2,i)}{s_0\left(s_1^2+1\right)\left(s_2^2+1\right)\left(s_2^2 s_3^2-1\right)} + \frac{\left(s_1^2 s_2^2 s_3^2 + s_3^2-1\right) s_2^2 g(5,i)}{\left(s_1^2+1\right)\left(s_2^2+1\right)\left(s_2^2 s_3^2-1\right)} -\\
&- \frac{s_0^2 s_3^2 s_2^2 g(4,i)}{\left(s_1^2+1\right)\left(s_2^2+1\right)\left(s_2^2 s_3^2-1\right)} + \frac{s_0 s_3\left(-s_1^2 s_2^2 + s_1^2 s_3^2 s_2^2-1\right) s_2 g(1,i)}{s_1\left(s_1^2+1\right)\left(s_2^2+1\right)\left(s_2^2 s_3^2-1\right)} -\\
&- \frac{\left(s_1^2-1\right)\left(s_2^2-1\right)\left(s_0^2 s_1^2 s_2^2+1\right)\left(s_3^2-1\right)\left(s_0^2 s_1^2 s_2^2 s_3^2-1\right) g(3,i)}{s_1^2\left(s_0^2 s_1^2-1\right)\left(s_1^2+1\right)\left(s_0^2 s_2^2-1\right)\left(s_2^2+1\right)\left(s_0^2 s_3^2-1\right)\left(s_2^2 s_3^2-1\right))},
\end{aligned}
$$

$$
\begin{aligned}
P_i^3 &= \frac{s_0^2 s_2^2 s_3^2 g(4,i)}{\left(s_1^2 s_2^2-1\right)\left(s_2^2+1\right)\left(s_3^2+1\right)} + \frac{s_1 s_2 s_3 g(2,i)}{s_0\left(s_1^2 s_2^2-1\right)\left(s_2^2+1\right)\left(s_3^2+1\right)} -\\
&- \frac{s_0 s_2\left(s_2^2 s_3^2 s_1^2 + s_1^2-1\right) s_3 g(1,i)}{s_1\left(s_1^2 s_2^2-1\right)\left(s_2^2+1\right)\left(s_3^2+1\right)} + \frac{\left(s_1^2 s_2^2 s_3^2 - s_2^2 s_3^2-1\right) g(5,i)}{\left(s_1^2 s_2^2-1\right)\left(s_2^2+1\right)\left(s_3^2+1\right)} -\\
&- \frac{\left(s_1^2-1\right)\left(s_2^2-1\right)\left(s_3^2-1\right)\left(s_0^2 s_1^2 s_2^2 s_3^2-1\right) g(3,i)}{s_1^2\left(s_0^2 s_2^2-1\right)\left(s_1^2 s_2^2-1\right)\left(s_2^2+1\right)\left(s_0^2 s_3^2-1\right)\left(s_3^2+1\right)}.
\end{aligned}
\tag{5.1}
$$

Here $g(1,i)$, $g(2,i)$, $g(3,i)$, $g(4,i)$ and $g(5,i)$ stand for the generators $G_i$, $G_i^{-1}$, $E_i$, $G_i^2$ and $1_i$ correspondingly. The first relation that can be readily obtained is 5-CB skein relation. It is a direct consequence of the completeness relation, (2.12), written in terms of the defined above generators

$$
G_i^3 = \alpha 1_i + \beta E_i + \gamma G_i + \delta G_i^{-1} + \mu G_i^2, \tag{5.2}
$$

where

$$
\begin{aligned}
\alpha &= -\frac{s_1\left(s_1^2 s_2^2 s_3^2 - s_2^2 s_3^2 + s_3^2 - 1\right)}{s_0^3 s_2 s_3^3},\\
\beta &= \frac{\left(s_1^2-1\right)\left(s_2^2-1\right)\left(s_0^2 s_1^2 s_2^2+1\right)\left(s_3^2-1\right)\left(s_0^2 s_1^2 s_2^2 s_3^2-1\right)}{s_0^5 s_1^3 s_2^3\left(s_0^2 s_2^2-1\right) s_3^3\left(s_0^2 s_3^2-1\right)},\\
\gamma &= \frac{s_1^2 s_3^2 s_2^4 + s_1^2 s_2^2 - s_1^2 s_3^2 s_2^2 + s_3^2 s_2^2 - s_2^2 + 1}{s_0^2 s_2^2 s_3^2},\\
\delta &= -\frac{s_1^2}{s_0^4 s_3^2},\\
\mu &= \frac{-s_2^2 s_1^2 + s_2^2 s_3^2 s_1^2 + s_1^2 - 1}{s_0 s_1 s_2 s_3}.
\end{aligned} \tag{5.3}
$$

Similarly, we get the delooping relation

$$
E_i G_i = G_i E_i = l^{-1} E_i, \tag{5.4}
$$

where

$$
l = s_0^3 s_1 s_2 s_3,
$$

which is consistent with eq. (4.5) and the idempotent relation

$$
E_i E_i = b E_i, \tag{5.5}
$$

where

$$
b = -\frac{\alpha}{\beta} + \frac{1}{\beta l^3} - \frac{\mu}{\beta l^2} - \frac{\gamma}{\beta l} - \frac{\delta l}{\beta}, \tag{5.6}
$$

which is consistent with the general expression, eq. (4.2).

As it follows from the explicit form of the trigonometric solution (2.13, 2.14), in 5-CB case the Yang Baxter equation contains 61 coefficients that accompany different powers of $e^{iu}$ and $e^{iv}$. All these coefficients have to be zero, which gives 61 relations for the trilinear combinations of the generators. We find it convenient to denote by $a_{i,j,k}(r,s,t)$ the element of the algebra $a_i[r]a_j[s]a_k[t]$, where $a_i[r]$ is $G_r, G_r^{-1}, E_r, G_r^2$ or $1_r$ according to whether $i = 1,2,3,4,5$, respectively.

Below the consequences of the skein arelation for the trilinear combinations $a_{i,j,k}[r,s,t]$ appearing in the YBE

$$
\begin{aligned}
&a_{4,2,3}(i,i\pm 1,i) = \alpha a_{5,3,3}(i,i\pm 1,i) + \beta a_{3,3,3}(i,i\pm 1,i) + \gamma a_{1,3,3}(i,i\pm 1,i) + \\
&+ \delta a_{2,3,3}(i,i\pm 1,i) + \mu a_{4,3,3}(i,i\pm 1,i), \\
&a_{2,1,3}(i,i\pm 1,i) = -\frac{(\beta l) a_{3,3,3}(i,i\pm 1,i)}{\delta} - \frac{\alpha a_{2,3,3}(i,i\pm 1,i)}{\delta} - \frac{\gamma a_{5,3,3}(i,i\pm 1,i)}{\delta} - \\
&- \frac{\mu a_{1,3,3}(i,i\pm 1,i)}{\delta} + \frac{a_{4,3,3}(i,i\pm 1,i)}{\delta}, \\
&a_{3,2,4}(i,i\pm 1,i) = \alpha a_{3,3,5}(i,i\pm 1,i) + \beta a_{3,3,3}(i,i\pm 1,i) + \gamma a_{3,3,1}(i,i\pm 1,i) + \\
&+ \delta a_{3,3,2}(i,i\pm 1,i) + \mu a_{3,3,4}(i,i\pm 1,i), \\
&a_{3,1,2}(i,i\pm 1,i) = -\frac{(\beta l) a_{3,3,3}(i,i\pm 1,i)}{\delta} - \frac{\alpha a_{3,3,2}(i,i\pm 1,i)}{\delta} - \frac{\gamma a_{3,3,5}(i,i\pm 1,i)}{\delta} - \\
&- \frac{\mu a_{3,3,1}(i,i\pm 1,i)}{\delta} + \frac{a_{3,3,4}(i,i\pm 1,i)}{\delta}, \\
&a_{4,5,4}(i,i\pm 1,i) = \left(\beta\mu + \frac{\beta}{l}\right) a_{5,5,3}(i,i\pm 1,i) + (\alpha + \gamma\mu) a_{5,5,1}(i,i\pm 1,i) + \\
&+ (\alpha\mu + \delta) a_{5,5,5}(i,i\pm 1,i) + \left(\gamma + \mu^2\right) a_{5,5,4}(i,i\pm 1,i) + \delta\mu a_{5,5,2}(i,i\pm 1,i).
\end{aligned}
\tag{5.7}
$$

Below are the consequences of the idempotent relation for the trilinear combinations $a_{i,j,k}(r,s,t)$ appearing in the YBE

$$
\begin{aligned}
a_{3,1,3}(i,i\pm 1,i) &= l a_{5,3,5}(i\pm 1,i,i\pm 1), \\
a_{3,2,3}(i,i\pm 1,i) &= \frac{a_{5,3,5}(i\pm 1,i,i\pm 1)}{l}, \\
a_{3,5,1}(i,i\pm 1,i) &= \frac{a_{5,3,5}(i\pm 1,i,i\pm 1)}{l}, \\
a_{1,5,3}(i,i\pm 1,i) &= \frac{a_{5,3,5}(i\pm 1,i,i\pm 1)}{l}, \\
a_{2,5,3}(i,i\pm 1,i) &= l a_{5,3,5}(i\pm 1,i,i\pm 1), \\
a_{3,5,2}(i,i\pm 1,i) &= l a_{5,3,5}(i\pm 1,i,i\pm 1)
\end{aligned}
\tag{5.8}
$$

We implement the relations eqs. (5.7, 5.8), together with other parameter-free relations, like braiding relations, Temperley–Lieb algebra relations, and their consequences.

Hence, below we list only the (parameter-free) relations which involve $a_4[r]$, i.e., $G_r^2$,

$$
\begin{aligned}
&a_{1,5,1}(i\pm1,i,i\pm1)=a_{5,4,5}(i,i\pm1,i),\\
&a_{4,1,1}(i\pm1,i,i\pm1)=a_{1,1,4}(i,i\pm1,i),\\
&a_{2,2,4}(i\pm1,i,i\pm1)=a_{4,2,2}(i,i\pm1,i),\\
&a_{2,4,1}(i\pm1,i,i\pm1)=a_{1,4,2}(i,i\pm1,i),\\
&a_{1,2,3}(i,i\pm1,i)=a_{4,3,3}(i,i\pm1,i),\\
&a_{3,2,1}(i,i\pm1,i)=a_{3,3,4}(i,i\pm1,i),\\
&a_{4,1,3}(i,i\pm1,i)=a_{1,3,3}(i,i\pm1,i),\\
&a_{3,1,4}(i,i\pm1,i)=a_{3,3,1}(i,i\pm1,i).
\end{aligned}
\tag{5.9}
$$

Notice that the first relation above is just the definition, $G_i^2$. While others come out as a result of YBE relations, or some simple algebra, e.g., in order to get $a_{4,1,3}(i,i\pm1,i)=a_{1,3,3}(i,i\pm1,i)$ we write $a_{4,1,3}(i,i\pm1,i)=G_i\,a_{1,1,3}(i,i\pm1,i)=G_i\,a_{3,3,5}(i\pm1,i,i\pm1)=a_{1,3,3}(i,i\pm1,i)\,1_{i\pm1}$.

After taking into account BMW′ relations together with eqs. (5.7-5.9), from the initial set of 61 relations that follow from YBE we are left with 37 relations. Not all these relations are independent, and give only 19 new relations, which are to be added to BMW′ plus five-block skein relation, in order to describe the full 5-CB algebra.

For 5-CB algebra, which is relevant for $SU(2)$ models, as it follows from eq. (3.3), we fix further

$$
s_0=q,\ \ s_1=q^2,\ \ s_2=q^3,\ \ s_3=q^4,
\tag{5.10}
$$

where, as described in the previous section, $q=e^{il}$ and $l=\frac{\pi}{k+2}$.

The obtained 19 5-CB relations for $SU(2)$ models have the following form (the relations are labeled by $m$):

$$
G_m=g_m(i,i+1,i)+\bar{g}_m(i+1,i,i+1)=0.
\tag{5.11}
$$

Here $g_m(i, i+1, i)$ stands for some linear combination of the elements $a_{i_n,j_n,k_n} = a_{i_n,j_n,k_n}(i, i+1, i)$ with some coefficients $k_{m,n}$ which depend on $q$,

$$g_m = \sum_n k_{m,n} a_{i_n,j_n,k_n} \tag{5.12}$$

and $\bar{g}_m(i+1, i, i+1)$ stands for some linear combination of elements $b_{i_n,j_n,k_n} = a_{i_n,j_n,k_n}(i+1, i, i+1)$ with $q$-dependent coefficients $r_{m,n}$,

$$\bar{g}_m = \sum_n r_{m,n} b_{i_n,j_n,k_n}.$$

It is assumed that the arguments $(i, i+1, i)$ or $(i+1, i, i+1)$ for $g$ or $\bar{g}$, correspondingly, refers to each element of the sum.

Suppressing the coefficients, we get that the elements $g_m$ can be schematically represented as follows

$$g_1 = a_{1,2,1} + a_{1,2,4} + a_{1,3,1} + a_{1,4,4} + a_{2,1,4} + a_{2,3,4} + a_{2,4,4} + a_{4,1,2} + a_{4,2,4} + a_{4,3,4}+$$
$$+ a_{4,4,1} + a_{5,1,2} + a_{5,1,3} + a_{5,1,4} + a_{5,2,1} + a_{5,2,3} + a_{5,2,4} + a_{5,3,1} + a_{5,3,2} + a_{5,3,4}+$$
$$+ a_{5,4,1} + a_{5,4,2} + a_{5,4,3} + a_{5,5,1} + a_{5,5,2} + a_{5,5,3}$$
$$g_2 = a_{1,3,1} + a_{1,4,4} + a_{2,3,4} + a_{2,4,2} + a_{2,4,4} + a_{4,1,2} + a_{4,1,4} + a_{4,2,4} + a_{4,3,4} + a_{5,1,2}+$$
$$+ a_{5,1,3} + a_{5,1,4} + a_{5,2,1} + a_{5,2,3} + a_{5,2,4} + a_{5,3,1} + a_{5,3,2} + a_{5,3,4} + a_{5,4,1} + a_{5,4,2}+$$
$$+ a_{5,4,3} + a_{5,5,1} + a_{5,5,2} + a_{5,5,3} + a_{5,5,4}$$
$$g_3 = a_{4,3,3} + a_{5,1,3} + a_{5,2,3} + a_{5,3,1} + a_{5,3,2} + a_{5,3,4} + a_{5,4,3} + a_{5,5,3}$$
$$g_4 = a_{1,2,4} + a_{1,3,1} + a_{1,4,4} + a_{2,1,4} + a_{2,3,4} + a_{4,1,2} + a_{4,2,1} + a_{4,3,4} + a_{4,4,1} + a_{5,1,2}+$$
$$+ a_{5,1,3} + a_{5,1,4} + a_{5,2,1} + a_{5,2,3} + a_{5,3,1} + a_{5,3,2} + a_{5,3,4} + a_{5,4,1} + a_{5,4,3} + a_{5,5,3}$$
$$g_5 = a_{3,4,1} + a_{5,5,3}$$
$$g_6 = a_{1,2,4} + a_{1,3,1} + a_{1,4,1} + a_{2,1,4} + a_{2,3,4} + a_{2,4,2} + a_{4,2,4} + a_{4,4,1} + a_{5,1,3} + a_{5,1,4}+$$
$$+ a_{5,2,3} + a_{5,2,4} + a_{5,3,1} + a_{5,3,2} + a_{5,3,4} + a_{5,4,1} + a_{5,4,2} + a_{5,4,3} + a_{5,5,1} + a_{5,5,2}+$$
$$+ a_{5,5,3} + a_{5,5,4}$$
$$g_7 = a_{1,3,4} + a_{2,3,4} + a_{4,3,4} + a_{5,3,4} + a_{5,4,3} + a_{5,5,3}$$
$$g_8 = a_{1,3,1} + a_{1,4,4} + a_{2,1,4} + a_{2,3,4} + a_{2,4,4} + a_{4,1,2} + a_{4,3,4} + a_{4,4,1} + a_{4,4,2} + a_{5,1,2} + a_{5,1,3}+$$
$$+ a_{5,1,4} + a_{5,2,1} + a_{5,2,3} + a_{5,2,4} + a_{5,3,1} + a_{5,3,2} + a_{5,3,4} + a_{5,4,1} + a_{5,4,2} + a_{5,4,3} + a_{5,5,3}$$
$$g_9 = a_{3,4,2} + a_{5,5,3}$$
$$g_{10} = a_{1,2,4} + a_{1,3,1} + a_{2,1,2} + a_{2,1,4} + a_{2,3,4} + a_{2,4,2} + a_{2,4,4} + a_{4,2,4} + a_{5,1,3} + a_{5,2,3} + a_{5,2,4}+$$
$$+ a_{5,3,1} + a_{5,3,2} + a_{5,3,4} + a_{5,4,2} + a_{5,4,3} + a_{5,5,1} + a_{5,5,2} + a_{5,5,3} + a_{5,5,4}$$
$$g_{11} = a_{2,4,3} + a_{5,1,3} + a_{5,2,3} + a_{5,3,1} + a_{5,3,2} + a_{5,3,4} + a_{5,4,3} + a_{5,5,3}$$
$$g_{12} = a_{1,3,1} + a_{2,3,4} + a_{4,3,1} + a_{5,1,3} + a_{5,3,1}$$
$$g_{13} = a_{4,4,3} + a_{5,1,3} + a_{5,2,3} + a_{5,3,1} + a_{5,3,2} + a_{5,3,4} + a_{5,4,3} + a_{5,5,3}$$
$$g_{14} = a_{3,4,4} + a_{5,5,3}$$
$$g_{15} = a_{1,3,1} + a_{2,3,4} + a_{4,3,2} + a_{4,3,4} + a_{5,2,3} + a_{5,3,2} + a_{5,3,4} + a_{5,4,3} + a_{5,5,3}$$

$g_{16} = a_{1,2,4} + a_{1,3,1} + a_{1,4,4} + a_{2,1,4} + a_{2,3,4} + a_{2,4,2} + a_{2,4,4} + a_{4,1,2} + a_{4,2,4} + a_{4,3,4} + a_{4,4,1} +$

$+ a_{4,4,4} + a_{5,1,2} + a_{5,1,3} + a_{5,1,4} + a_{5,2,1} + a_{5,2,3} + a_{5,2,4} + a_{5,3,1} + a_{5,3,2} + a_{5,3,4} + a_{5,4,1} +$

$+ a_{5,4,2} + a_{5,4,3} + a_{5,5,1} + a_{5,5,2} + a_{5,5,3} + a_{5,5,4}$

$g_{17} = a_{3,4,3} + a_{5,5,3}$

$g_{18} = a_{1,4,3} + a_{5,1,3} + a_{5,2,3} + a_{5,3,1} + a_{5,3,2} + a_{5,3,4} + a_{5,4,3} + a_{5,5,3}$

$g_{19} = a_{3,3,4} + a_{5,5,3}$

The elements $\bar{g}_m$ are listed below in the schematic form:

$\bar{g}_1 = b_{1,2,1} + b_{1,3,1} + b_{1,4,4} + b_{2,1,4} + b_{4,1,2} + b_{4,2,1} + b_{4,2,4} + b_{4,3,2} + b_{4,3,4} + b_{4,4,1} + b_{4,4,2} +$

$+ b_{5,1,2} + b_{5,1,3} + b_{5,1,4} + b_{5,2,1} + b_{5,2,3} + b_{5,2,4} + b_{5,3,1} + b_{5,3,2} + b_{5,3,4} + b_{5,4,1} + b_{5,4,2} +$

$+ b_{5,4,3} + b_{5,5,1} + b_{5,5,2} + b_{5,5,3}$

$\bar{g}_2 = b_{1,3,1} + b_{2,1,4} + b_{2,4,2} + b_{4,1,4} + b_{4,2,4} + b_{4,3,2} + b_{4,3,4} + b_{4,4,1} + b_{4,4,2} + b_{5,1,3} + b_{5,1,4} +$

$+ b_{5,2,3} + b_{5,2,4} + b_{5,3,1} + b_{5,3,2} + b_{5,3,4} + b_{5,4,1} + b_{5,4,2} + b_{5,4,3} + b_{5,5,1} + b_{5,5,2} + b_{5,5,3} + b_{5,5,4}$

$\bar{g}_3 = b_{3,3,4} + b_{5,5,3}$

$\bar{g}_4 = b_{1,2,4} + b_{1,3,1} + b_{1,4,4} + b_{2,1,4} + b_{4,1,2} + b_{4,2,1} + b_{4,3,2} + b_{4,3,4} + b_{4,4,1} + b_{5,1,2} + b_{5,1,3} +$

$+ b_{5,1,4} + b_{5,2,1} + b_{5,2,3} + b_{5,3,1} + b_{5,3,2} + b_{5,3,4} + b_{5,4,1} + b_{5,4,3} + b_{5,5,3}$

$\bar{g}_5 = b_{1,4,3} + b_{5,1,3} + b_{5,2,3} + b_{5,3,1} + b_{5,3,2} + b_{5,3,4} + b_{5,4,3} + b_{5,5,3}$

$\bar{g}_6 = b_{1,3,1} + b_{1,4,1} + b_{1,4,4} + b_{2,4,2} + b_{4,1,2} + b_{4,2,1} + b_{4,2,4} + b_{4,3,2} + b_{5,1,2} + b_{5,1,3} + b_{5,2,1} +$

$+ b_{5,2,3} + b_{5,2,4} + b_{5,3,1} + b_{5,3,2} + b_{5,3,4} + b_{5,4,2} + b_{5,4,3} + b_{5,5,1} + b_{5,5,2} + b_{5,5,3} + b_{5,5,4}$

$\bar{g}_7 = b_{4,3,1} + b_{4,3,2} + b_{4,3,4} + b_{5,1,3} + b_{5,2,3} + b_{5,3,1} + b_{5,3,2} + b_{5,3,4} + b_{5,4,3} + b_{5,5,3}$

$\bar{g}_8 = b_{1,3,1} + b_{1,4,4} + b_{2,1,4} + b_{2,4,4} + b_{4,1,2} + b_{4,3,2} + b_{4,3,4} + b_{4,4,1} + b_{4,4,2} + b_{5,1,2} + b_{5,1,3} +$

$+ b_{5,1,4} + b_{5,2,1} + b_{5,2,3} + b_{5,2,4} + b_{5,3,1} + b_{5,3,2} + b_{5,3,4} + b_{5,4,1} + b_{5,4,2} + b_{5,4,3} + b_{5,5,3}$

$\bar{g}_9 = b_{2,4,3} + b_{5,1,3} + b_{5,2,3} + b_{5,3,1} + b_{5,3,2} + b_{5,3,4} + b_{5,4,3} + b_{5,5,3}$

$\bar{g}_{10} = b_{1,3,1} + b_{2,1,2} + b_{2,4,2} + b_{4,1,2} + b_{4,2,1} + b_{4,2,4} + b_{4,3,2} + b_{4,4,2} + b_{5,1,2} + b_{5,1,3} + b_{5,2,1} +$

$+ b_{5,2,3} + b_{5,2,4} + b_{5,3,1} + b_{5,3,2} + b_{5,3,4} + b_{5,4,2} + b_{5,4,3} + b_{5,5,1} + b_{5,5,2} + b_{5,5,3} + b_{5,5,4}$

$\bar{g}_{11} = b_{3,4,2} + b_{5,5,3}$

$\bar{g}_{12} = b_{1,3,1} + b_{1,3,4} + b_{4,3,2} + b_{5,2,3} + b_{5,3,2}$

$$\bar{g}_{13} = b_{3,4,4} + b_{5,5,3}$$

$$\bar{g}_{14} = b_{4,4,3} + b_{5,1,3} + b_{5,2,3} + b_{5,3,1} + b_{5,3,2} + b_{5,3,4} + b_{5,4,3} + b_{5,5,3}$$

$$\bar{g}_{15} = b_{1,3,1} + b_{2,3,4} + b_{4,3,2} + b_{4,3,4} + b_{5,2,3} + b_{5,3,2} + b_{5,3,4} + b_{5,4,3} + b_{5,5,3}$$

$$\bar{g}_{16} = b_{1,3,1} + b_{1,4,4} + b_{2,1,4} + b_{2,4,2} + b_{4,1,2} + b_{4,2,1} + b_{4,2,4} + b_{4,3,2} + b_{4,3,4} +$$
$$+ b_{4,4,1} + b_{4,4,2} + b_{4,4,4} + b_{5,1,2} + b_{5,1,3} + b_{5,1,4} + b_{5,2,1} + b_{5,2,3} + b_{5,2,4} +$$
$$+ b_{5,3,1} + b_{5,3,2} + b_{5,3,4} + b_{5,4,1} + b_{5,4,2} + b_{5,4,3} + b_{5,5,1} + b_{5,5,2} + b_{5,5,3} + b_{5,5,4}$$

$$\bar{g}_{17} = b_{3,4,3} + b_{5,5,3}$$

$$\bar{g}_{18} = b_{3,4,1} + b_{5,5,3}$$

$$\bar{g}_{19} = b_{4,3,3} + b_{5,1,3} + b_{5,2,3} + b_{5,3,1} + b_{5,3,2} + b_{5,3,4} + b_{5,4,3} + b_{5,5,3} \tag{5.13}$$

A sample few 5-CB algebra relations for $su(2)$ models, eq.(5.11), which are sufficiently short, are listed below explicitly:

$$G_3 = \frac{\left(q^{18} - q^{14} - q^{12} + q^8 - 1\right) a_{5,1,3}}{q^8} + \frac{\left(q^{18} - q^{14} - q^{12} + q^8 - 1\right) a_{5,3,2}}{q^8} -$$
$$- \frac{\left(q^{18} - q^{14} - q^{12} + q^8 - 1\right) a_{5,2,3}}{q^8} - \frac{\left(q^{18} - q^{14} - q^{12} + q^8 - 1\right) a_{5,3,1}}{q^8} +$$
$$+ \frac{\left(q^{28} - q^{24} - q^{22} - q^{20} + 2q^{18} + q^{16} - q^{12} - 2q^{10} + q^6 + q^4 - 1\right) a_{5,5,3}}{q^8} -$$
$$- \frac{\left(q^{28} - q^{24} - q^{22} - q^{20} + 2q^{18} + q^{16} - q^{12} - 2q^{10} + q^6 + q^4 - 1\right) b_{5,5,3}}{q^8} +$$
$$+ b_{3,3,4} - a_{4,3,3} - a_{5,3,4} + a_{5,4,3}$$

$$G_5 = \frac{\left(q^{24} - q^{18} + q^{14} + q^{10} - q^6 + 1\right) a_{5,5,3}}{q^4} - a_{3,4,1} + q^4 \left(-b_{5,3,4}\right) + q^4 b_{5,4,3} -$$
$$-(q-1)(q+1)\left(q^2+1\right)\left(q^2-q+1\right)\left(q^2+q+1\right)\left(q^{10}-q^8+q^6-q^2+1\right) q^2 b_{5,2,3}+$$
$$+(q-1)(q+1)\left(q^2+1\right)\left(q^2-q+1\right)\left(q^2+q+1\right)\left(q^{10}-q^8+q^6-q^2+1\right) q^2 b_{5,3,2}+$$
$$+\frac{(q-1)(q+1)\left(q^2+1\right)\left(q^2-q+1\right)\left(q^2+q+1\right)\left(q^{10}-q^8+q^4-q^2+1\right) b_{5,1,3}}{q^4}-$$
$$-\frac{(q-1)(q+1)\left(q^2+1\right)\left(q^2-q+1\right)\left(q^2+q+1\right)\left(q^{10}-q^8+q^4-q^2+1\right) b_{5,3,1}}{q^4}-$$
$$-\frac{\left(q^{24} - q^{18} + q^{14} + q^{10} - q^6 + 1\right) b_{5,5,3}}{q^4} + b_{1,4,3}$$

$$G_7 = -\frac{q^8 a_{4,3,4}}{q^{18}-q^{14}-q^{12}+q^8-1} + \frac{q^8 b_{4,3,4}}{q^{18}-q^{14}-q^{12}+q^8-1} - a_{1,3,4} + a_{2,3,4} +$$
$$+ b_{4,3,1} - b_{4,3,2} + \frac{\left(q^{28}-q^{24}-q^{22}-q^{20}+2q^{18}+q^{16}-q^{12}-2q^{10}+q^6+q^4-1\right) a_{5,4,3}}{q^{18}-q^{14}-q^{12}+q^8-1}+$$
$$+\frac{\left(q^{28}-q^{24}-q^{22}-q^{20}+2q^{18}+q^{16}-q^{12}-2q^{10}+q^6+q^4-1\right)^2 a_{5,5,3}}{q^8\left(q^{18}-q^{14}-q^{12}+q^8-1\right)}-$$
$$-\frac{\left(q^{28}-q^{24}-q^{22}-q^{20}+2q^{18}+q^{16}-q^{12}-2q^{10}+q^6+q^4-1\right) a_{5,3,4}}{q^{18}-q^{14}-q^{12}+q^8-1}+$$
$$+\frac{\left(q^{28}-q^{24}-q^{22}-q^{20}+2q^{18}+q^{16}-q^{12}-2q^{10}+q^6+q^4-1\right) b_{5,2,3}}{q^8}+$$
$$+\frac{\left(q^{28}-q^{24}-q^{22}-q^{20}+2q^{18}+q^{16}-q^{12}-2q^{10}+q^6+q^4-1\right) b_{5,3,1}}{q^8}-$$
$$-\frac{\left(q^{28}-q^{24}-q^{22}-q^{20}+2q^{18}+q^{16}-q^{12}-2q^{10}+q^6+q^4-1\right) b_{5,1,3}}{q^8}-$$
$$-\frac{\left(q^{28}-q^{24}-q^{22}-q^{20}+2q^{18}+q^{16}-q^{12}-2q^{10}+q^6+q^4-1\right) b_{5,3,2}}{q^8}+$$
$$+\frac{\left(q^{28}-q^{24}-q^{22}-q^{20}+2q^{18}+q^{16}-q^{12}-2q^{10}+q^6+q^4-1\right) b_{5,3,4}}{q^{18}-q^{14}-q^{12}+q^8-1}-$$
$$-\frac{\left(q^{28}-q^{24}-q^{22}-q^{20}+2q^{18}+q^{16}-q^{12}-2q^{10}+q^6+q^4-1\right) b_{5,4,3}}{q^{18}-q^{14}-q^{12}+q^8-1}-$$
$$-\frac{\left(q^{28}-q^{24}-q^{22}-q^{20}+2q^{18}+q^{16}-q^{12}-2q^{10}+q^6+q^4-1\right)^2 b_{5,5,3}}{q^8\left(q^{18}-q^{14}-q^{12}+q^8-1\right)}$$

$$G_9 = -(q^{42} - q^{38} - q^{36} - q^{34} + 2q^{32} + 2q^{30} - 2q^{26} - 4q^{24} + q^{22} + 2q^{20} + 2q^{18} - 2q^{14} -$$
$$- q^{12} - q^{10} + q^8 + q^6 - 1)a_{5,5,3}/q^{28} - a_{3,4,2} + \frac{\left(q^{18} - q^{14} - q^{12} + q^8 - 1\right) b_{5,3,4}}{q^{16}} -$$
$$- \frac{\left(q^{18} - q^{14} - q^{12} + q^8 - 1\right) b_{5,4,3}}{q^{16}} +$$
$$+ \frac{(q-1)(q+1)\left(q^2+1\right)\left(q^2-q+1\right)\left(q^2+q+1\right)\left(q^{10} - q^8 + q^4 - q^2 + 1\right) b_{5,3,2}}{q^{12}} -$$
$$- \frac{(q-1)(q+1)\left(q^2+1\right)\left(q^2-q+1\right)\left(q^2+q+1\right)\left(q^{10} - q^8 + q^4 - q^2 + 1\right) b_{5,2,3}}{q^{12}} +$$
$$+ (q-1)(q+1)\left(q^2+1\right)\left(q^2-q+1\right)\left(q^2+q+1\right) \times$$
$$\times \frac{\left(q^{36} - 2q^{34} + q^{32} + 2q^{26} - 3q^{24} + q^{18} + q^{16} - q^{14} - 2q^{10} + q^8 + q^6 - q^4 + q^2 - 1\right) b_{5,3,1}}{q^{24}} -$$
$$- (q-1)(q+1)\left(q^2+1\right)\left(q^2-q+1\right)\left(q^2+q+1\right) \times$$
$$\times \frac{\left(q^{36} - 2q^{34} + q^{32} + 2q^{26} - 3q^{24} + q^{18} + q^{16} - q^{14} - 2q^{10} + q^8 + q^6 - q^4 + q^2 - 1\right) b_{5,1,3}}{q^{24}} +$$
$$+ (q^{42} - q^{38} - q^{36} - q^{34} + 2q^{32} + 2q^{30} - 2q^{26} - 4q^{24} + q^{22} + 2q^{20} + 2q^{18} - 2q^{14} - q^{12} -$$
$$- q^{10} + q^8 + q^6 - 1)b_{5,5,3}/q^{28} + b_{2,4,3}$$

$$G_{12} = -\frac{\left(q^{18} - q^{14} - q^{12} + q^8 - 1\right) a_{1,3,1}}{q^8} + \frac{\left(q^{18} - q^{14} - q^{12} + q^8 - 1\right) b_{1,3,1}}{q^8} - a_{4,3,1} +$$
$$+ \frac{\left(q^{28} - q^{24} - q^{22} - q^{20} + 2q^{18} + q^{16} - q^{12} - 2q^{10} + q^6 + q^4 - 1\right) a_{5,1,3}}{q^8} + b_{1,3,4} -$$
$$- \frac{\left(q^{28} - q^{24} - q^{22} - q^{20} + 2q^{18} + q^{16} - q^{12} - 2q^{10} + q^6 + q^4 - 1\right) a_{5,3,1}}{q^8} - a_{2,3,4} +$$
$$+ \frac{\left(q^{28} - q^{24} - q^{22} - q^{20} + 2q^{18} + q^{16} - q^{12} - 2q^{10} + q^6 + q^4 - 1\right) b_{5,3,2}}{q^8} + b_{4,3,2} -$$
$$- \frac{\left(q^{28} - q^{24} - q^{22} - q^{20} + 2q^{18} + q^{16} - q^{12} - 2q^{10} + q^6 + q^4 - 1\right) b_{5,2,3}}{q^8}$$

$$G_{17} = \frac{\left(q^{20} - q^{15} + q^{10} - q^5 + 1\right)\left(q^{20} + q^{15} + q^{10} + q^5 + 1\right) a_{5,5,3}}{q^{20}} - a_{3,4,3} -$$
$$- \frac{\left(q^{20} - q^{15} + q^{10} - q^5 + 1\right)\left(q^{20} + q^{15} + q^{10} + q^5 + 1\right) b_{5,5,3}}{q^{20}} + b_{3,4,3}$$

$$
\begin{aligned}
G_{18} &= q^4 a_{5,3,4} - q^4 a_{5,4,3} - a_{1,4,3} + b_{3,4,1} + \\
&+ (q-1)(q+1)\left(q^2+1\right)\left(q^2-q+1\right)\left(q^2+q+1\right)\left(q^{10}-q^8+q^6-q^2+1\right) q^2 a_{5,2,3} - \\
&- (q-1)(q+1)\left(q^2+1\right)\left(q^2-q+1\right)\left(q^2+q+1\right)\left(q^{10}-q^8+q^6-q^2+1\right) q^2 a_{5,3,2} + \\
&+ \frac{(q-1)(q+1)\left(q^2+1\right)\left(q^2-q+1\right)\left(q^2+q+1\right)\left(q^{10}-q^8+q^4-q^2+1\right) a_{5,3,1}}{q^4} - \\
&- \frac{(q-1)(q+1)\left(q^2+1\right)\left(q^2-q+1\right)\left(q^2+q+1\right)\left(q^{10}-q^8+q^4-q^2+1\right) a_{5,1,3}}{q^4} + \\
&+ \frac{\left(q^{24}-q^{18}+q^{14}+q^{10}-q^6+1\right) a_{5,5,3}}{q^4} - \frac{\left(q^{24}-q^{18}+q^{14}+q^{10}-q^6+1\right) b_{5,5,3}}{q^4}
\end{aligned}
$$

We checked these relations for the $SU(2)$ Boltzmann weights described in section (3), for general values of the parameter $q$, using various heights, and we found that they are all obeyed.

The full relations, which are too long to list here, can be found in the attached Mathematica file.

## 6. Conclusions.

In this work we discussed the $n$–CB (Conformal Braiding) algebra, where $n$ is the number of blocks, i.e., the degree of the polynomial equation obeyed by the Boltzmann weights. We concentrated on the algebra for more than 5 blocks, $n \geq 5$. We used the fused $m \times m$ $SU(2)$ model of Date et al. [17], as a testing ground for the $n$–CB algebra, and showed that it obeys a version of the BMW algebra [15, 16]. We described explicitly the 5–CB algebra specialized to the $SU(2)$ model, using the expansion of the Yang–Baxter relation. We propose that this method can be used to get the $n$–CB algebra for any $n$ and any model, assuming that the baxterization ansatz holds.

Our results are useful in knot theory. As was shown by Wadati et al. [5], any solvable IRF model gives a link invariant, which is expressed in terms of the Boltzmann weights. See also [19]. The calculation of the knot invariants is complicated, when using only the Boltzmann weights. However, the $n$–CB algebra can be used to simplify it considerably, enabling the calculation of the link invariant for any knot, using only simple several explicit calculations for any model, directly from the Wadati et al. Markov trace [5].

Acknowledgements: We thank J.R. Li, R. Tessler and H. Wenzl for useful discussions and I. Deichaite for remarks on the manuscript. D.G. thanks the ICTP, Trieste, for the hospitality where part of this work was done. The research of V.B. was supported by the RFBR under grant 18-02-01131.